\documentclass[12pt]{article}

\usepackage{amssymb}
\usepackage{amsmath}
\usepackage{graphicx}

\catcode`@=11
\renewcommand{\section}{\@startsection{section}{1}{0pt}{\medskipamount}
{\medskipamount}{\large\bf}}
\numberwithin{equation}{section}
\catcode`@=12

\def\beq{\begin{eqnarray}}    
\def\eeq{\end{eqnarray}}      

\def\ln{\,\mbox{ln}\,}                  

\def\im{\textrm{i}}

\def\diff{\textrm{d}}
\def\sfrac#1#2{{\textstyle\frac{#1}{#2}}}
\def\={\ =\ }
\def\und{\qquad\textrm{and}\qquad}

\def\al{\alpha}

\def\de{\delta}

\def\vp{\varepsilon}

\begin{document}

\begin{titlepage}
\setcounter{page}{0}
\begin{flushright}
ITP--UH--06/11
\end{flushright}

\vskip 2.0cm

\begin{center}

{\LARGE\bf  
Is soft breaking of BRST symmetry consistent?}

\vspace{18mm}

{\Large
Peter Lavrov$\,{}^{\dagger}$, \
Olaf Lechtenfeld$\,{}^{\times}$ \ and \ 
Alexander Reshetnyak$\,{}^\ast$
}

\vspace{8mm}

\noindent ${}^\dagger${\em
Tomsk State Pedagogical University,\\  
Kievskaya St.\ 60, 634061 Tomsk, Russia} 

\vspace{4mm}

\noindent ${}^\times${\em
Institut f\"ur Theoretische Physik, Leibniz Universit\"at Hannover,\\ 
Appelstrasse 2, 30167 Hannover, Germany}

\vspace{4mm}

$^\ast${\em
Institute of Strength Physics and Material Science, \\ 
Akademicheskii av.\ 2/4, 634021 Tomsk, Russia}

\vspace{18mm}

\begin{abstract}
\noindent 
A definition of soft breaking of BRST symmetry in the field-antifield formalism 
is proposed, valid for general gauge theories and arbitrary gauge fixing.
The Ward identities for the generating functionals of Green's functions are 
derived, and their gauge dependence is investigated. We discuss  the Gribov-Zwanziger action for the one-parameter family of $R_\xi$ gauges. It is argued that gauge theories with a soft breaking of BRST symmetry are inconsistent.
\end{abstract}

\end{center}

\vfill
\noindent{\sl Emails:} \ lavrov@tspu.edu.ru, lechtenf@itp.uni-hannover.de, reshet@ispms.tsc.ru


\end{titlepage}


\section{Introduction}

\noindent
It is well known that BRST symmetry~\cite{brst}, as a global
fermionic remnant of gauge invariance, plays a fundamental role 
in quantum field theory, because all fundamental forces existing 
in Nature can be described in terms of gauge theories~\cite{books}.

Recently, in a series of papers~\cite{Sorellas} based on Zwanziger's
action~\cite{Zwanziger1,Zwanziger2}, a breakdown of BRST symmetry 
in Yang-Mills theories has been considered from a new point of view. 
This breakdown is related to attempts to take into account the Gribov 
horizon~\cite{Gribov}, which restricts the domain of integration in 
the functional integral presenting the Green's functions of the given 
gauge theory. Effectively this restriction can be implemented by a 
particular addition to the standard Faddeev-Popov~(FP) action. However,
this addition is not invariant under the original BRST transformations.

We remark that until now all investigations
\cite{Sorellas,Zwanziger1,Zwanziger2} of the Gribov horizon in 
Yang-Mills theories have been performed in the Landau gauge only. 
Yet there is a great freedom in the choice of admissible gauges, 
and it is well known that the Green's functions depend on the gauge. 
Of course, this dependence is structured, as it must cancel in physical
combinations such as the S-matrix. Modern proofs (see, e.g.,~\cite{LT}) 
of the gauge independence of the S-matrix for Yang-Mills theories are 
based on BRST symmetry. Any violation of BRST invariance may, therefore, 
spell doom for the consistency of the gauge theory. Thus, any claim of
a breakdown of BRST symmetry warrants serious investigation.

Modern models of the fundamental forces make use of gauge theories 
more general than Yang-Mills theory. Luckily, the concept of BRST
invariance generalizes to supergravity, theories with an open gauge 
algebra and reducible gauge theories, to name a few.
The present paper formulates the soft breaking of BRST symmetry for 
general gauge theories in the field-antifield formalism~\cite{BV}. We then 
investigate the gauge dependence of the Green's functions for arbitrary 
gauge models with softly broken BRST symmetry.

The paper is organized as follows. In Section~2, our definition of the
soft breaking of BRST symmetry is given in the field-antifield formalism.
Using a suitable regularization scheme, Section~3 derives the Ward 
identities for the customary generating functionals of Green's functions.
In Section~4 we investigate the dependence of these functionals on an 
arbitrary gauge, for general gauge theories. A discussion of the
Gribov-Zwanziger action for the one-parameter family of $R_\xi$ gauges is 
considered in Section~5. Finally, Section 6 gives concluding remarks.

We employ the condensed notation of DeWitt~\cite{DeWitt}.
Derivatives with respect to sources and antifields are taken from the left,
while those with respect to fields are taken from the right.
Left derivatives with respect to fields are labeled by a subscript~$l$. 
The Grassmann parity of any quantity $A$ is denoted by $\varepsilon (A)$.

\section{Soft breaking of BRST symmetry}

\noindent
Our starting point is a theory of  gauge fields $A^i$, $i=1,2,\ldots,n$, 
with $\varepsilon(A^i)=\varepsilon_i$, described by an initial
action $S_0=S_0(A)$ invariant under the gauge transformations
\begin{eqnarray}
\label{GIClassA}
\delta A^i= R^i_{\alpha}(A)\xi^{\alpha}
\qquad\textrm{hence}\qquad
S_{0,i}(A) R^i_{\alpha}(A)=0 \qquad\textrm{for}\quad
\alpha=1,2,\ldots,m\ ,\quad 0<m<n\ ,
\end{eqnarray}
parametrized by $m$ arbitrary functions $\xi^{\alpha}$ of the space-time
coordinates, with $\varepsilon(\xi^{\alpha}) =\varepsilon_{\alpha}$.
Here, $S_{0,i}\equiv\delta S_0/\delta A^i$, and $R^i_{\alpha}(A)$
are the generators of the gauge transformations, with
$\vp(R^i_\al)=\vp_i{+}\vp_\al$.
We shall not restrict ourselves to some special type of initial gauge theory;
it may belong to open gauge theories and/or reducible gauge theories. The
type of initial gauge theory defines the structure of configuration
space $\{\Phi^A\}$ in the field-antifield formalism~\cite{BV},
\begin{eqnarray}
\label{ConfSp}
\Phi\ \equiv\ \{\Phi^A\}\=\{A^i,\ldots\} \qquad\textrm{with}\qquad
\varepsilon(\Phi^A)=\varepsilon_A\ ,
\end{eqnarray}
where the dots indicate the full set of ghost and antighost fields,
auxiliary fields and so on. In what follows we do not need to
describe the exact structure of the full configuration space. 
To each field $\Phi^A$ of this total configuration space, one introduces 
the corresponding antifield~$\Phi^*_A$, hence
\beq \label{antiSpaceRBV} 
\Phi^*\ \equiv\ \{\Phi^*_A\} \= \{A^*_i,\ldots\}\ , \eeq
with statistics opposite to that of the
corresponding fields $\Phi^A$, i.e.~$\vp(\Phi^*_A)=\vp_A{+}1$.

On the total space of the fields $\Phi^A$ and the antifields $\Phi^*_A$, 
one defines a bosonic functional ${\bar S}={\bar S}(\Phi,\Phi^*)$
satisfying the master equation
\beq \label{MastEBV} 
\sfrac {1}{2} ({\bar S},{\bar
S})\=\im\hbar\,{\Delta}{\bar S} \eeq
with the boundary condition
\beq \label{BoundCon} {\bar S}|_{\Phi^* = \hbar = 0}\= S_0(A)\ . \eeq
In (\ref{MastEBV}) we used the notation of the antibracket
\beq \label{DefAB}
(F, G)\ \equiv \ \frac{\delta
F}{\delta\Phi^A}\,\frac{\delta G}{\delta\Phi^*_A}\ -\
(F\leftrightarrow G)\;
(-1)^{\left[\vp(F)+1\right]\cdot\left[\vp(G)+1\right]} \eeq 
and of the nilpotent  operator 
\beq \label{DeltaBV}
\Delta\ \equiv\ (-1)^{\vp_A} \frac{\delta_{\it l}}
{\delta\Phi^A}\;\frac{\delta} {\delta\Phi^*_A} \qquad\textrm{with}\qquad
{\Delta}^2=0 \quad\textrm{and}\quad \vp (\Delta)=1\ . \eeq
We  assume that formal manipulations with the operator $\Delta$ can be
supported by a suitable regularization scheme. This is a nontrivial
requirement, since the operator (\ref{DeltaBV}) is not well-defined
on local functionals. The reason is that for any local functional
$F$, one finds that $\Delta F\sim\delta(0)$. 
The usual way to deal with this problem is to use
dimensional regularization~\cite{Leib}, which equates $\delta(0)$ 
to zero. In this paper, we shall imply such a type of regularization, 
so that the master equation is reduced to the classical master equation
\beq \label{ClMastEBV}
({\bar S},{\bar S})\=0\ .
\eeq
Using the action ${\bar S}$ and a  fermionic gauge fixing functional
$\Psi = \Psi(\Phi)$, one can construct the  non-degenerate action
$S_{ext}$ by the rule
\beq \label{ExtActBV} 
S_{ext}(\Phi, \Phi^*) \= {\bar S}\big(
\Phi,\,\Phi^* + \sfrac{\de\Psi}{\de\Phi} \big)\ . \eeq
This action satisfies the classical master equation 
\beq \label{ClMastEBVExt}
(S_{ext}, S_{ext})\=0 \eeq 
and is used to construct the generating functional of Green's
functions in the field-antifield formalism~\cite{BV}.

Inspired by \cite{Zwanziger1,Zwanziger2}, we modify the
action $S_{ext}$ by adding a functional $M=M(\Phi,\Phi^*)$,
defining the full action $S$ as 
\beq \label{Sfull}
S\=S_{ext}+M\ . \eeq 
We shall speak of a soft breaking of BRST
symmetry in the field-antifield formalism if the condition 
\beq \label{SoftBrC} (M,M)\=0 \eeq 
is fulfilled. Therefore, the basic
classical equation of our approach to the soft breaking of BRST symmetry
reads 
\beq \label{CBasEq} \sfrac{1}{2}(S,S)\=(S,M)\ . \eeq 
If the soft breaking of BRST symmetry originates from a modification of
the integration measure, then $M$ will be a functional of the field variables
$\Phi^A$ only, i.e.~$M=M(\Phi)$. In this case, the condition
(\ref{SoftBrC}) is automatically valid. In fact, this is exactly the
situation for Yang-Mills theory in Landau gauge,
when one takes into account the Gribov horizon~\cite{Zwanziger1,Zwanziger2}. 
We do not restrict ourselves to this special case and consider the 
more general situation of $M=M(\Phi,\Phi^*)$.

It is interesting to note that the right-hand side of the basic classical 
equation~(\ref{CBasEq}) can be presented in the form 
\beq (S,M)\={\hat s}M \ ,\eeq 
where ${\hat s}$ denotes the Slavnov-Taylor operator defined by the rule 
\beq {\hat s}\=(S_{ext},\bullet)\ . \eeq 
Due to (\ref{ClMastEBVExt}) this operator is nilpotent, 
\beq {\hat s}^2\=0\ , \eeq 
and we find that
\beq {\hat s}\,(S,S)\=0\ . \eeq
On this level we formally meet the same relation as for general
gauge theories {\it without\/} a soft breaking of BRST symmetry.

\section{Generating functionals and Ward identities}

\noindent
Let us consider some quantum consequences of the
classical equations (\ref{ClMastEBVExt}), (\ref{SoftBrC}) and~(\ref{CBasEq}). 
To this end we introduce the generating functional
of Green's functions, 
\beq \label{ZBV}
Z(J,\Phi^*)\=\int\!D\Phi\ \exp \Big\{\frac{\im}{\hbar}
\big(S(\Phi,\Phi^*)+ J_A\Phi^A\big)\Big\}\ , \eeq 
where $S(\Phi,\Phi^*)$ satisfies the basic classical equation
(\ref{CBasEq}) and has the form~(\ref{Sfull}). Furthermore, 
$J_A$ are the usual external sources for the fields~$\Phi^A$. 
The Grassmann parities of these sources are defined in a
natural way, \ $\vp(J_A) = \vp_A$.

 From (\ref{ClMastEBVExt}) it follows that
\beq \nonumber 
0 \= \int\!D\Phi\ 
\frac{\delta S_{ext}}{\delta\Phi^A}\frac{\delta S_{ext}}{\delta\Phi^*_A }\;
\exp \Big\{\frac{\im}{\hbar} \big(S_{ext}+ M+J_A\Phi^A\big)\Big\}\ . \eeq
Performing the usual
manipulations with the functional integral we arrive at the following
identity for the generating functional $Z$, 
\beq \label{WIZBV}
\frac{\hbar}{\im}\Big(J_A+M_{A}\big(\sfrac{\hbar}{\im}\sfrac{\delta}{\delta
J},\Phi^*\big)\Big)\frac{\delta Z(J,\Phi^*)}{\delta\Phi^*_A}\ -\ J_A
M^{A*}\big(\sfrac{\hbar}{\im}\sfrac{\delta}{\delta
J},\Phi^*\big)Z(J,\Phi^*)\=0\ . \eeq 
Here the notations 
\beq \nonumber M_{A}\big(\sfrac{\hbar}{\im}\sfrac{\delta}{\delta
J},\Phi^*\big)\equiv \frac{\delta M(\Phi,\Phi^*)}{\delta
\Phi^A}\Big|_{\Phi\rightarrow \frac{\hbar}{\im}\frac{\delta}{\delta J}} 
\und
M^{A*}\big(\sfrac{\hbar}{\im}\sfrac{\delta}{\delta J},\Phi^*\big)\equiv
\frac{\delta M(\Phi,\Phi^*)}{\delta\Phi^*_A}\Big|_{\Phi\rightarrow
\frac{\hbar}{\im}\frac{\delta}{\delta J}} \eeq 
have been used. In case of $M=0$, the identity  (\ref{WIZBV}) is
reduced to the usual Ward identity for the generating functional of Green's
functions in the field-antifield formalism. Hence, we refer to~(\ref{WIZBV}) 
as the Ward identity for~$Z$
in a gauge theory with softly broken BRST symmetry.

Introducing the generating functional of connected Green's
functions, 
\beq W(J,\Phi^*) \= -\im\hbar \ln Z(J,\Phi^*)\ ,\eeq
the identity (\ref{WIZBV}) can be rewritten as 
\beq \label{WIWBV}
\Big(J_A+M_{A}\big(\sfrac{\delta W}{\delta J}+
\sfrac{\hbar}{\im}\sfrac{\delta}{\delta J},\Phi^*\big)\Big)\frac{\delta
W(J,\Phi^*)}{\delta\Phi^*_A}\ -\ J_A M^{A*}\big(\sfrac{\delta W}{\delta
J}+\sfrac{\hbar}{\im}\sfrac{\delta}{\delta J},\Phi^*\big)=0\ . \eeq

The generating functional of the vertex functions or effective action 
is obtained by Legendre transforming~$W$,
\beq
\label{EA}
\Gamma (\Phi,\,\Phi^*) \= W(J,\Phi^*) - J_A\Phi^A
\qquad\textrm{where}\quad
\Phi^A = \frac{\delta W}{\delta J_A}
\quad\textrm{and}\quad
\frac{\delta\Gamma}{\delta\Phi^A}=-J_A\ .
\eeq
Taking into account the equality 
\beq \nonumber \frac{\delta
\Gamma}{\delta \Phi^*_A}\=\frac{\delta W}{\delta\Phi^*_A}\ , \eeq 
we can rewrite the identity (\ref{WIWBV}) in terms of~$\Gamma$ as
\beq \label{WIGammaBV} 
\sfrac{1}{2}(\Gamma,\Gamma) \=
\frac{\delta\Gamma}{\delta\Phi^A}{\widehat M}^{A*}
+{\widehat M}_{A}\frac{\delta \Gamma}{\delta
\Phi^*_A}\ .
\eeq
Here, we have used the notation 
\beq {\widehat M}_{A}\ \equiv\ 
\frac{\delta M(\Phi,\Phi^*)}{\delta\Phi^A}\Big|_{\Phi\to\widehat\Phi}
\und
{\widehat M}^{A*}\ \equiv\ 
\frac{\delta M(\Phi,\Phi^*)}{\delta\Phi^*_A}\Big|_{\Phi\to\widehat\Phi}
\eeq
where
\beq {\widehat\Phi}^A\=\Phi^A+\im\hbar\,(\Gamma^{''-1})^{AB}
\frac{\delta_l}{\delta\Phi^B} \eeq 
and the matrix $(\Gamma^{''-1})$ is inverse to the matrix
$\Gamma^{''}$ with elements
\beq (\Gamma^{''})_{AB}\=\frac{\delta_l}{\delta\Phi^A}
\Big(\frac{\delta\Gamma}{\delta\Phi^B}\Big)\ ,\qquad\textrm{i.e.}\quad
(\Gamma^{''-1})^{AC}(\Gamma^{''})_{CB}=\delta^A_{\ B}\ . \eeq 
Again, in the case $M=0$ the identity (\ref{WIGammaBV}) coincides with the Ward
identity for the effective action in the field-antifield formalism. Note
that the identity (\ref{WIGammaBV}) is compatible with the basic
classical equation (\ref{CBasEq}), since $\hbar\to0$ yields
$\Gamma=S$, ${\widehat M}=M$, and (\ref{WIGammaBV}) is reduced to~(\ref{CBasEq}). 

But this is not the end of the story, because on the classical level we also
have the restriction~(\ref{SoftBrC}), i.e. $(M,M)=0$, which on the quantum
level generates an additional identity. To derive it we consider
a direct consequence of~(\ref{SoftBrC}),
\beq \label{extra}
0 \=\int\!D\Phi\ \frac{\delta
M}{\delta\Phi^A}\frac{\delta M}{\delta\Phi^*_A }\ \exp
\Big\{\frac{\im}{\hbar} \big(S_{ext}+ M+J_A\Phi^A\big)\Big\}\ .
\eeq
Omitting the details of the functional-integral manipulations,
we can rewrite~(\ref{extra}) as 
\beq \label{SoftBrCq}  {\widehat M}_{A}{\widehat M}^{A*} \=0\ . \eeq

In addition, the relations $\Delta M=0$ and $\Delta S=0$ yield 
quantum consequences. Indeed, from the evident equality
\beq \nonumber
0\=\int\!D\Phi\ (-1)^{\varepsilon_A}\frac{\delta_l}{\delta\Phi^A}
\Big[\frac{\delta M}{\delta\Phi^*_A}\;\exp \Big\{\frac{\im}{\hbar}
\big(S+J_A\Phi^A\big)\Big\}\Big] \eeq
it follows that
\beq \nonumber
0\=\int\!D\Phi\ \Big[\Delta M+\frac{\im}{\hbar}\Big(J_A+\frac{\delta
S}{\delta\Phi^A}\Big)\frac{\delta M}{\delta\Phi^*_A}\Big]\;\exp
\Big\{\frac{\im}{\hbar} \big(S+J_A\Phi^A\big)\Big\}\ , \eeq
which produces the identity 
\beq \label{AddIdM} 
\Big(\frac{\delta \Gamma}{\delta\Phi^A}- 
{\widehat S}_{A}\Big){\widehat M}^{A*}\=0 \ ,\eeq 
where 
\beq \label{hatdefs}
{\widehat S}_A\equiv S_A({\widehat\Phi},\Phi^*)= \frac{\delta
S(\Phi,\Phi^*)}{\delta\Phi^A}\Big|_{\Phi\rightarrow{\widehat\Phi}} 
\und {\widehat\Phi}^A= 
\Phi^A+\im\hbar\,(\Gamma^{''-1})^{AB}\frac{\delta_l}{\delta\Phi^B}\ .
\eeq 
In turn starting with the equality 
\beq \nonumber 
0\=\int\!D\Phi\ (-1)^{\varepsilon_A}\frac{\delta_l}{\delta\Phi^A}
\Big[\frac{\delta S}{\delta\Phi^*_A}\;\exp \Big\{\frac{\im}{\hbar}
\big(S+J_A\Phi^A\big)\Big\}\Big]\ , 
\eeq 
we have 
\beq \nonumber 
0\=\int\!D\Phi\ \Big[\Delta S+\frac{\im}{\hbar}\Big(J_A+\frac{\delta
S}{\delta\Phi^A}\Big)\frac{\delta S}{\delta\Phi^*_A}\Big]\;\exp
\Big\{\frac{\im}{\hbar} \big(S+J_A\Phi^A\big)\Big\}\ , 
\eeq
which gives us
\beq \label{AddIdEqDS} 
\Big(\frac{\delta\Gamma}{\delta\Phi^A}-{\widehat S}_{A}\Big)
\frac{\delta \Gamma}{\delta \Phi^*_A}\=0
\eeq
as the quantum version of the equality $\Delta S=0$. 

Finally, it should be noted that from the equality
\beq \nonumber
0\= \int\!D\Phi\ \frac{\delta}{\delta\Phi^A}
\Big[\exp \Big\{\frac{\im}{\hbar} \big(S+J_A\Phi^A\big)\Big\}\Big]
\eeq
we can derive in the usual manner the relation 
\beq \label{AddIdEqS} 
\frac{\delta \Gamma}{\delta\Phi^A}\={\widehat S}_{A} \ ,\eeq
which is nothing but the other representation of equation  for~$\Gamma$ (see (\ref{EA})). 
It implies the relations (\ref{AddIdM}) and (\ref{AddIdEqDS}).

Analogously, starting with the identity
\beq
\frac{\delta F(\Phi)}{\delta\Phi^*_A}\equiv 0
\eeq
for an arbitrary functional $F(\Phi)$, we get
\beq 
0 \= \int\!D\Phi\ \frac{\delta F(\Phi)}{\delta\Phi^*_A}\;
\exp \Big\{\frac{\im}{\hbar} \big(S+ J_A\Phi^A\big)\Big\}
\eeq
and therefore
\beq \label{MGamma}
F(\widehat{\Phi})\frac{\delta \Gamma}{\delta\Phi^{*}_A} \=
F(\widehat{\Phi})\,S^{A*}(\widehat{\Phi},\Phi^*)\ . \eeq
Since the functional $F(\Phi)$ was arbitrary, we also have the relation
\beq 
\frac{\delta \Gamma}{\delta\Phi^{*}_A} \= \widehat{S}^{A*}
\eeq
with 
\beq \nonumber 
{\widehat S}^{A*}\equiv S^{A*}({\widehat\Phi},\Phi^*)= \frac{\delta
S(\Phi,\Phi^*)}{\delta\Phi_A^*}\Big|_{\Phi\rightarrow{\widehat\Phi}}\ .
\eeq
Clearly, in the $\hbar\to0$ limit ${\widehat M}=M$, and the
identity (\ref{SoftBrCq}) is reduced to (\ref{SoftBrC}). Therefore,
we have a full set of equalities which describe on the classical and
quantum level general gauge theories with a soft breaking of BRST
symmetry in arbitrary gauges within the field-antifield formalism.

\section{Gauge dependence}

\noindent
We turn to a discussion of gauge dependence of the generating functionals
$Z$, $W$ and $\Gamma$ for general gauge theories with a soft breaking of BRST
symmetry as defined in the previous section. 
The derivation of this dependence is based on the fact
that any variation of the gauge-fixing functional,
$\Psi(\Phi)\rightarrow\Psi(\Phi)+\delta\Psi(\Phi)$, leads to
a variation of the action $S_{ext}$ (\ref{ExtActBV}) and the functional
$Z$~\cite{VLT}. The variation of $S_{ext}$ can be presented in the form 
\beq
\label{varSext} \delta S_{ext}\=\frac{\delta
\delta\Psi}{\delta\Phi^A}\,\frac{\delta S_{ext}}{\delta\Phi^*_A}
\eeq 
or as 
\beq \label{varSext1} 
\delta S_{ext}\=-(S_{ext},\delta\Psi)\=-{\hat s}\,\delta\Psi\ . \eeq
We also allow the functional $M$ to be gauge dependent, with
$\delta M(\Phi,\Phi^*)$ being its variation simultaneous to the
variation~$\delta\Psi$ of the gauge-fixing functional.
 From (\ref{ZBV}), (\ref{varSext}) and the variation of $M$ 
we obtain the gauge variation of~$Z$, 
\beq \label{varZ} 
\delta Z(J,\Phi^*)\=\frac{\im}{\hbar}\int\!D\Phi\ \Big( \frac{\delta
\delta\Psi}{\delta\Phi^A}\frac{\delta S_{ext}}{\delta\Phi^*_A}+\delta
M\Big)\; \exp \Big\{\frac{\im}{\hbar}
\big(S(\Phi,\Phi^*)+ J_A\Phi^A\big)\Big\}\ . \eeq
With the help of
\beq \label{AuxId} \nonumber
0&=&\int\!D\Phi\ \frac{\delta_l}{\delta
\Phi^A}\Big[\delta\Psi\;\frac{\delta S_{ext}}{\delta\Phi^*_A}\;\exp
\Big\{\frac{\im}{\hbar} \big(S(\Phi,\Phi^*)+
J_A\Phi^A\big)\Big\}\Big] \\ \nonumber
&=&\int\!D\Phi\ \Big[\frac{\delta\delta\Psi
}{\delta\Phi^A}\,\frac{\delta
S_{ext}}{\delta\Phi^*_A}-\frac{\im}{\hbar}\Big(J_A+\frac{\delta
S}{\delta \Phi^A}\Big)\frac{\delta
S_{ext}}{\delta\Phi^*_A}\,\delta\Psi\Big]\exp \Big\{\frac{\im}{\hbar}
\big(S(\Phi,\Phi^*)+ J_A\Phi^A\big)\Big\}\ , \eeq 
where $\Delta S_{ext}=0$ was used, and the relation
\beq \nonumber 
\frac{\delta S}{\delta\Phi^A}\,\frac{\delta
S_{ext}}{\delta\Phi^*_A}\= \frac{\delta M}{\delta\Phi^A}\,\frac{\delta
S}{\delta\Phi^*_A}\ , \eeq 
we can rewrite (\ref{varZ}) as
\beq \nonumber 
\delta Z(J,\Phi^*)&=&\frac{\im}{\hbar}\Big[\Big(J_A+M_{A}
\big(\sfrac{\hbar}{\im}\sfrac{\delta}{\delta
J},\Phi^*\big)\Big)\frac{\delta}{\delta\Phi^*_A}\,
\delta\Psi\big(\sfrac{\hbar}{\im
}\sfrac{\delta}{\delta J}\big) 
\ -\ \frac{\im}{\hbar}J_A M^{A*}
\big(\sfrac{\hbar}{\im}\sfrac{\delta}{\delta J},\Phi^*\big)\,\delta\Psi
\big(\sfrac{\hbar}{\im}\sfrac{\delta}{\delta J}\big) \\ \label{varZ1}
&&\quad +\ \delta M\big(\sfrac{\hbar}{\im}\sfrac{\delta}{\delta
J},\Phi^*\big)\Big]Z(J,\Phi^*) \\[4pt] \label{deltaZfin1}
&=&\frac{\im}{\hbar}\Big[\hat{q}\;\delta\Psi
\big(\sfrac{\hbar}{\im}\sfrac{\delta}{\delta J}\big)\ +\
\delta M\big(\sfrac{\hbar}{\im}\sfrac{\delta}{\delta
J},\Phi^*\big) \Big] Z(J,\Phi^*)\ ,
\eeq
where we have abbreviated the first line by introducing 
the nilpotent fermionic operator
\beq \label{q} \hat{q}
\=\Big(J_A+M_{A}\big(\sfrac{\hbar}{\im}\sfrac{\delta}{\delta
J},\Phi^*\big)\Big)\frac{\delta }{\delta\Phi^*_A}\ -\
\frac{\im}{\hbar}J_A
M^{A*}\big(\sfrac{\hbar}{\im}\sfrac{\delta}{\delta
J},\Phi^*\big)\ . \eeq
Its nilpotency, $\hat{q}^2 = 0$, is proved in the Appendix.

The corresponding variation of the generating functional of connected 
Green's functions takes the form
\beq \label{varW}
\delta W(J,\Phi^*)\= \sfrac{\hbar}{\im}Z^{-1}\delta Z\=
{\hat Q}\; \delta\Psi
\big(\sfrac{\delta W}{\delta J}+\sfrac{\hbar}{\im}\sfrac{\delta}{\delta J}\big)
\ +\ \delta M\big(\sfrac{\delta W}{\delta
J}+\sfrac{\hbar}{\im}\sfrac{\delta}{\delta
J},\Phi^*\big) \ ,
\eeq
where the fermionic operator ${\hat Q}$ is unitarily related to $\hat{q}$,
\beq \label{Qexpl} 
{\hat Q}\=\exp\big\{{-}\sfrac{\im}{\hbar}W\big\}\;{\hat
q}\;\exp\big\{\sfrac{\im}{\hbar}W\big\} \=
\Big(J_A+M_{A}\big(\sfrac{\delta
W}{\delta J}+\sfrac{\hbar}{\im}\sfrac{\delta}{\delta
J},\Phi^*\big)\Big)\frac{\delta }{\delta\Phi^*_A}\ ,
\eeq
with the help of the Ward identity (\ref{WIWBV}).
Note that all terms in $\hat{Q}$ contain an antifield derivative. 
 From its construction, ${\hat Q}$ is nilpotent as well, i.e.~${\hat Q}^2=0$.

Let us proceed to the gauge variation of the effective action.  
We firstly note that $\delta \Gamma=\delta W$. 
Secondly, we observe that the definitions~(\ref{EA}) and the 
Ward identity~(\ref{WIGammaBV}) imply that
\beq \label{dphistar}
{\frac{\delta}{\delta\Phi^*}}\Big|_{J}\=
{\frac{\delta}{\delta\Phi^*}}\Big|_{\Phi} +\ \frac{\delta
\Phi}{\delta\Phi^*}\,{\frac{\delta_{\it l}}{\delta\Phi}}\Big|_{\Phi^*}\ .
\eeq 
Next, differentiating the Ward identities~(\ref{WIZBV}) with respect 
to the sources $J$, then rewriting
these relations for the functional $W$ and transforming the latter
with allowance for (\ref{EA}) and~(\ref{WIGammaBV}), we arrive at
\beq 
{\hat Q}\,\Phi^{A}\big|_{J}&=& \Big({\widehat M}^{A*}-\frac {\delta
\Gamma}{\delta\Phi^*_A}\Big)(-1)^{\varepsilon_A} \nonumber \\
&& +\ \frac{\im}{\hbar}\Big(\Phi^{A}{\widehat M}_{B}
\frac {\delta \Gamma}{\delta \Phi^{*}_{B}}(-1)^{\varepsilon_A}-
{\widehat M}_{B} \frac {\delta \Gamma}{\delta
\Phi^{*}_{B}}\Phi^{A}\Big)\nonumber \\  \label{QPhi} 
&& +\ \frac{\im}{\hbar}\Big(\Phi^{A}\frac {\delta \Gamma}{\delta
\Phi^{B}}{\widehat M}^{B^*} (-1)^{\varepsilon_A}-
\frac {\delta \Gamma}{\delta \Phi^{B}}{\widehat M}^{B^*} \Phi^{A}\Big)\ . %
\eeq
 From (\ref{varW})--(\ref{QPhi}) we can represent the gauge variation of
the effective action in the following form, 
\beq
\delta\Gamma \= {\hat s}_q\,\langle\delta\Psi\rangle\ +\ 
\langle\delta M\rangle \label{varGamma}\ , \eeq
where the operator ${\hat s}_q$ is given by
\beq 
{\hat s}_q &=& -(\Gamma,\bullet)\ +\ {\widehat M}_{A}
\frac{\delta}{\delta \Phi^{*}_{A}}\ +\ (-1)^{\vp_A}
{\widehat M}^{A*} \frac{\delta_l}{\delta \Phi^A} \nonumber \\
&& -\ \frac{\im}{\hbar}\Big({\widehat M}_B \frac
{\delta\Gamma}{\delta\Phi^*_B}\Phi^A-
(-1)^{\varepsilon_A}\Phi^A {\widehat M}_B \frac
{\delta\Gamma}{\delta\Phi^*_B}\Big) 
\frac{\delta_{\it l}}{\delta\Phi^A}\nonumber \\
&& -\ \frac{\im}{\hbar}\Big(\frac{\delta\Gamma}{\delta
\Phi^B}{\widehat M}^{B*} \Phi^A-
(-1)^{\varepsilon_A}\Phi^A \frac{\delta\Gamma}{\delta
\Phi^B}{\widehat M}^{B*} \Big) 
\frac{\delta_{\it l}}{\delta\Phi^A} \label{hats} 
\eeq
and we introduced the notation 
\beq
\langle\delta\Psi\rangle\=\delta\Psi({\widehat\Phi})\cdot 1 \und
\langle\delta M\rangle\=\delta M({\widehat \Phi},\Phi^*)\cdot 1 \ . 
\eeq
Because ${\hat s_q}$ is related to ${\hat Q}$ via a Legendre
transformation (which is a change of variables),  
it must be nilpotent as well, ${\hat s}^2_q=0$.

Another extremely useful representation for $\delta\Gamma$ is
obtained by a slightly different rewriting of~(\ref{QPhi}) as follows, 
\beq
\frac {\delta \Gamma}{\delta\Phi^{*}_{A}}- {\widehat M}^{A*}
&=& -(-1)^{\vp_A\vp_B}\Big({\widehat M}_B -\frac {\delta
\Gamma}{\delta\Phi^B} \Big)(\Gamma^{''-1})^{AC}
\frac{\delta_{\it l}}{\delta\Phi^C}\frac{\delta\Gamma}{\delta\Phi^*_B}
\nonumber \\
&& +\ \frac{\im}{\hbar}\Big(\Phi^{A}{\widehat M}_B
\frac{\delta\Gamma}{\delta\Phi^*_B}-
(-1)^{\varepsilon_A}{\widehat M}_B \frac{\delta
\Gamma}{\delta\Phi^*_B}\Phi^A\Big)\nonumber \\
&& +\ \frac{\im}{\hbar}\Big(\Phi^A \frac{\delta
\Gamma}{\delta\Phi^B}{\widehat M}^{B*} -
(-1)^{\varepsilon_A}\frac{\delta\Gamma}{\delta
\Phi^B}{\widehat M}^{B*} \Phi^A\Big)\ .\label{Wdiff}
\eeq
As the result we obtain our final expression for the gauge variation
of the effective action,
\beq
\delta\Gamma \= \frac{\delta\Gamma}{\delta\Phi^A}
{\widehat F}^A\,\langle\delta\Psi\rangle\ -\ 
{\widehat M}_A{\widehat F}^A \langle\delta\Psi\rangle\ +\
\langle\delta M\rangle\ , \label{varGammaF}
\eeq
with the operator definition
\beq  {\widehat F}^A \= 
-\frac{\delta}{\delta\Phi^*_A}\ -\ (-1)^{\vp_B(\vp_A+1)}
(\Gamma^{''-1})^{BC}\Big(\frac{\delta_{\it l}}{\delta\Phi^C}\frac
{\delta \Gamma}{\delta\Phi^{*}_{A}}\Big)\frac{\delta_{\it l}
}{\delta\Phi^B}\ . \label{FAdef}
\eeq

We see from (\ref{varGammaF}) that on shell the effective action 
is generally gauge dependent since
\beq \frac{\delta\Gamma}{\delta\Phi^A}=0 \qquad
\longrightarrow\qquad \delta\Gamma\neq 0\ . 
\eeq 
This negates a consistent formulation of a soft breaking of BRST
symmetry within the field-antifield formalism, 
unless perhaps the two last terms in~(\ref{varGammaF}) cancel each other,
\beq \label{BasRest} 
\langle\delta M\rangle\={\widehat M}_A{\widehat F}^A \langle\delta\Psi\rangle\ . 
\eeq 
This is a severe restriction on the BRST-breaking functional~$M$
for the effective action to be gauge independent on-shell.
The same statement is valid for physical S-matrix.
In fact, (\ref{BasRest})~fixes the gauge variation of $M=M(\Phi,\Phi^*)$
under a change of the gauge-fixing functional~$\Psi$ to be
\beq \label{BEqvM}
\delta M\=\frac{\delta M}{\delta\Phi^A}\,{\widehat F}_0^A\,\delta\Psi \eeq
where (see (\ref{Sfull}) and (\ref{CBasEq}))
\beq\label{F_0}
{\widehat F}_0^A\=-(-1)^{\vp_B(\vp_A+1) } (S^{''-1})^{BC}
\Big(\frac{\delta_{\it
l}}{\delta\Phi^C}\frac {\delta
S}{\delta\Phi^{*}_{A}}\Big)
\frac{\delta_{\it l} }{\delta\Phi^B}\ .
\eeq
%


\section{Gribov-Zwanziger action in a one-parameter family of gauges}
\noindent
In this section we shall apply our above-described general consideration 
of a soft BRST breaking to the important case of Yang-Mills theories, 
since those had been the subject of recent investigations~\cite{Sorellas}.
The initial classical action $S_0$ of Yang-Mills fields $A^a_{\mu}(x)$,
which take values in the adjoint representation of~$su(N)$ so that,
$a=1,\ldots,N^2{-}1$, has the standard form
\beq
S_0(A) \= -\sfrac14\int\!\diff^D x\ F_{\mu\nu}^{a}F^{\mu\nu{}a}
\qquad\textrm{with}\quad
F^a_{\mu\nu}\=\partial_{\mu}A^a_{\nu}-\partial_{\nu}A^a_{\mu}+
f^{abc}A^b_{\mu}A^c_{\nu}\ , \label{clYM}
\eeq
where $\mu,\nu=0,1,\ldots,D{-}1$, the Minkowski space has signature
$(-,+,\ldots,+)$, and $f^{abc}$ denote the (totally antisymmetric) structure 
constants of the Lie algebra~$su(N)$.
The action (\ref{clYM}) is invariant under the gauge transformations
\beq \delta A^a_{\mu}\=D^{ab}_{\mu}\xi^b \qquad\textrm{with}\quad
D^{ab}_{\mu}\=\delta^{ab}\partial_{\mu}+f^{acb}A^c_{\mu}\ .\eeq
The field configuration space of Yang-Mills theory, 
\beq \{\Phi^A\}\=\{A^a_{\mu}, B^a, C^a, {\bar C}^a\}
\qquad\textrm{with}\quad
\varepsilon(C^a)=\varepsilon(\bar C^a)=1\ ,\quad
\varepsilon(A^a_\mu)=\varepsilon(B^a)=0\ , \eeq
includes the (scalar) Faddeev-Popov ghost and antighost fields 
$C^a$ and ${\bar C}^a$, respectively, as well as the
Nakanishi-Lautrup auxiliary fields $B^a$. 
The corresponding set of antifields is
\beq \{\Phi^*_A\} \=\{A^{*a\mu}, B^{*a}, C^{*a}, {\bar C}^{*a}\}
\qquad\textrm{with}\quad
\varepsilon(A^{*a\mu})=\varepsilon(B^{*a})=1\ ,\quad
\varepsilon(C^{*a})= \varepsilon({\bar C}^{*a})=0\ . \eeq
A solution to the classical master equation (\ref{ClMastEBV}) 
can be presented in the form
\beq \bar{S}(\Phi,\Phi^*) \= S_0(A)\ +\ A^{*a\mu}D^{ab}_{\mu}C^b\ +\
\sfrac{1}{2}C^{*a}f^{abc}C^bC^c\ +\ \bar{C}{}^{*a}B^a\label{SbosGZ}\ .\eeq
The gauge-fixing functional can be chosen as
\beq \Psi(\Phi)\={\bar C}^a\chi^a(A,B) \eeq
with free bosonic functions~$\chi^a$,
so that the non-degenerate action $S_{ext}$~(\ref{ExtActBV}) becomes
\beq \nonumber 
S_{ext}(\Phi,\Phi^*)&=& S_0(A) + \Big(A^{*a\mu}+{\bar
C}^c\frac{\delta\chi^c}{\delta A^a_{\mu}}\Big)D^{ab}_{\mu}C^b +
\sfrac{1}{2}C^{*a}f^{abc}C^bC^c +
\big(\bar{C}{}^{*a}+\chi^a\big)B^a \\
&=&S_{FP}(\Phi)\ +\ A^{*a\mu}D^{ab}_{\mu}C^b\ +\
\sfrac{1}{2}C^{*a}f^{abc}C^bC^c\ +\ \bar{C}{}^{*a}B^a\ ,
\label{ExtYMBV}\eeq
where $S_{FP}(\Phi)$ is the Faddeev-Popov action
\beq\label{FPact}
S_{FP}(\Phi)\=S_0(A)\ +\ {\bar C}^a K^{ab}C^b\ +\ \chi^a B^a
\qquad\textrm{with}\qquad
K^{ab}=\frac{\delta\chi^a}{\delta A^c_{\mu}}D^{cb}_{\mu}\ .
\eeq
The actions (\ref{FPact}) and (\ref{ExtYMBV}) are invariant
under the BRST transformation
\beq\label{BRSTtr}
\delta_B A_{\mu}^{a} = D^{ab}_{\mu}C^b\theta\ ,\quad 
\delta_B \bar{C}{}^a = B^a\theta\ ,\quad 
\delta_B B^a = 0\ ,\quad 
\delta_B C^a = \sfrac12 f^{abc}C^bC^c\theta 
\label{BRSTGZred} \eeq
where $\theta$ is a constant Grassmann parameter.

In \cite{Zwanziger1,Zwanziger2} it has been shown that the
Gribov horizon \cite{Gribov} in Yang-Mills theory (\ref{clYM})
in the Landau gauge,
\beq \label{Landau}
\chi^a(A,B)\=\partial^{\mu}A_{\mu}^a
\qquad\longrightarrow\qquad
K^{ab}=\partial^\mu D_\mu^{ab}\ , 
\eeq
can be taken in to account by adding to the Faddeev-Popov action (\ref{FPact})
the non-local functional~\footnote{
    The choice of \cite{Sorellas} agrees with ours after Wick rotation,
    integrating out auxiliary fields and renaming $\gamma^4\to\gamma^2$.}
\beq \label{FuncM}
M(A)\=\gamma^2\,\big(f^{abc}A^b_{\mu}(K^{-1})^{ad}f^{dec}
A^{e\mu}\ +\ D(N^2{-}1)\big)\ , \eeq
where $K^{-1}$ is the matrix inverse to the Faddeev-Popov operator
$K^{ab}$ in~(\ref{Landau}).
The so-called thermodynamic or Gribov parameter~$\gamma$ is
determined in a self-consistent way by the gap equation
\cite{Zwanziger1,Zwanziger2}
\begin{equation}
\frac{\partial \mathcal{E}_{vac}}{\partial \gamma}=0\ ,\label{gapeq}
\end{equation}
where $\mathcal{E}_{vac}$ is the vacuum energy given by
\begin{equation}
\exp\Big\{\frac{\im}{\hbar}\mathcal{E}_{vac}\Big\}\=\int\!D\Phi\
\exp\Big\{\frac{\im}{\hbar}S_{GZ}(\Phi)\Big\} \label{vacEn}
\end{equation}
pertaining to the Gribov-Zwanziger action \cite{Sorellas}
\beq \label{GZact} S_{GZ}(\Phi)\=S_{FP}(\Phi)\ +\ M(A)\ . \eeq
Note that the functional $M(A)$ in~(\ref{FuncM}) is not invariant
under the BRST transformation (\ref{BRSTtr}) but trivially
satisfies the condition~(\ref{SoftBrC}) of soft BRST breaking 
because of its independence on antifields.

The Gribov-Zwanziger action was intensively investigated in
a series of papers~\cite{Sorellas} where various quantum properties
of gauge models with this action have been studied. We stress however
that it was impossible in principle to establish the gauge independence
of physical quantities in these theories because they were
formulated in the Landau gauge~(\ref{Landau}) only.
Here, we are going to clarify this crucial issue.

To this end, we discuss the Gribov-Zwanziger action (\ref{GZact})
for the one-parameter family of $R_\xi$ gauges, 
\beq 
\chi^a(A,B,\xi) 
\=\partial^{\mu}A_{\mu}^a\ +\ \sfrac{\xi}{2}B^a
\label{gcYM} \eeq
with a real parameter $\xi$ interpolating between the Landau gauge ($\xi{=}0$)
and the Feynman gauge~($\xi{=}1$). 
The Faddeev-Popov action is then written as
\beq
S_{FP}({\Phi,\xi})\=
S_0(A)\ +\ {\bar C}^a \partial^\mu D_\mu^{ab}C^b\ +\
(\partial^\mu A_\mu^a)B^a \ +\ \sfrac{\xi}{2} B^a B^a\ .\eeq
The Faddeev-Popov operator $K^{ab}$ is obviously independent of~$\xi$, 
but the functional $M$ must be modified away from $\xi{=}0$, already because
$K^{ab}$ ceases to be hermitian~\cite{SS}. Although a suitable functional
$M(A,B,\xi)$ is not known, we assume its existence with
\beq
\label{lMxi}
\lim_{\xi\rightarrow 0} M(A,B,\xi)\=M(A)
\eeq
where $M(A)$ is given by~(\ref{FuncM}).
Now we propose the Gribov-Zwanziger action for Yang-Mills theories 
(\ref{clYM}) in the $R_\xi$ gauge family~(\ref{gcYM}) as
\beq S_{GZ}(\Phi,\xi)\=S_{FP}(\Phi,\xi)\ +\ M(A,B,\xi)\ . \eeq
Because the BRST transformation (\ref{BRSTtr}) 
does not depend on the gauge fixing,
from~(\ref{lMxi}) by continuity we can conclude that
\beq
\delta_B M(A,B,\xi)\neq 0 \qquad\longrightarrow\qquad
\delta_BS_{GZ}(\Phi,\xi)\neq 0\ .
\eeq
Let us recall our consistency condition~(\ref{BEqvM}), which takes the form
\beq \delta
M(A,B,\xi) \ \buildrel{!}\over{=}\ 
\sfrac12\frac{\delta M(A,B,\xi)}{\delta\Phi^A}\,
{\widehat F}_0^A\,\bar{C}^aB^a\,\delta\xi \ .\eeq
Since the right-hand side necessarily depends on the ghost, antighost 
or auxiliary fields, it cannot match the left-hand side for our choice of $M$. 
Therefore, soft breaking of BRST symmetry is not consistent in $R_{\xi}$ gauges. 

\newpage

\section{Conclusions}

\noindent
We have proposed a definition of soft breaking of BRST symmetry
in the field-antifield formalism. To this end, a
`breaking functional'~$M$ had to be added to the gauge-fixed action
$S_{ext}$. The latter is constructed from an arbitrary classical
gauge-invariant action $S_0$ with the rules of the field-antifield method.
In terms of the functional $M$, the soft breaking of BRST symmetry was
defined by the analog of the classical master equation $(M,M)=0$. We have
derived all Ward identities for the generating functional of Green's functions,
of connected Green's functions and of vertex functions, denoted by $Z$,
$W$ and $\Gamma$, respectively. These identities were employed to
investigate the gauge dependence of those functionals. It was shown that
$\Gamma$ as well as the S-matrix are on-shell gauge dependent in general.
We discussed  the Gribov-Zwanziger action 
for the one-parameter family of $R_\xi$ gauges. Already in this simple case, 
the functional~$\Gamma$ turned out to depend on the gauge even on shell.
We are forced to conclude that a consistent quantization of gauge
theories with a soft breaking of BRST symmetry does not exist.

\section*{Acknowledgments}
\noindent
The authors thank I.L.~Buchbinder and I.V.~Tyutin for useful
discussions. P.L.~is grateful to the Institute for Theoretical
Physics of Leibniz Universit\"at Hannover for warm hospitality.
We are grateful to various authors of \cite{Sorellas} for a critical 
assessment of earlier versions.
This work was partially supported by the joint DFG grant 436 RUS 113/669/4.
The work of P.L.~is also supported by the LRSS grant 3558.2010.2, the
RFBR-Ukraine grant 11-02-90445 and the RFBR grant 09-02-00078.


\appendix
\section*{Appendix: \ Proof of nilpotency for $\hat{q}$}
\renewcommand{\theequation}{A.\arabic{equation}}
\setcounter{equation}{0}

\noindent
For simplicity of writing  let us abbreviate
\beq
M_{A}\big(\sfrac{\hbar}{\im}\sfrac{\delta}{\delta
J},\Phi^*\big)=:M_{A} \und
M^{A*}\big(\sfrac{\hbar}{\im}\sfrac{\delta}{\delta J},\Phi^*\big)=:M^{A*}\ .
\eeq
The square of $\hat{q}$ may be directly presented as a sum of four
operators,
\beq \hat{q}^2 & = &
\Big[\Big(J_A+M_{A}\Big)\frac{\delta }{\delta\Phi^*_A}-
\frac{\im}{\hbar}J_A
M^{A*}\Big]^2\ \equiv\ \sum_{i=1}^4 D_i\nonumber \\
& = &  \Big(J_A+M_{A}\Big)\frac{\delta
}{\delta\Phi^*_A}\Big(J_B+M_{B}\Big)\frac{\delta }{\delta\Phi^*_B}
\ -\  \frac{\im}{\hbar}
\Big(J_A+M_{A}\Big)\frac{\delta }{\delta\Phi^*_A}\;J_B
M^{B*}\nonumber \\
&&-\  \frac{\im}{\hbar} J_B
M^{B*}\Big(J_A+M_{A}\Big)\frac{\delta }{\delta\Phi^*_A}
+  \Big(\frac{\im}{\hbar}\Big)^2J_A
M^{A*}\; J_B
M^{B*} .\label{q2} \eeq
After rearranging the antifield derivatives,
the four summands in (\ref{q2}) take the form
\beq D_1 &=& (-1)^{\vp_A+1}
\big(J_A+M_{A}\big)\big(J_B+M_{B}\big)\frac{\delta
}{\delta\Phi^*_B}\frac{\delta
}{\delta\Phi^*_A}\  +\ \big(J_A+M_{A}\big) \frac{\delta
M_{B}}{\delta\Phi^*_A}
\frac{\delta}{\delta\Phi^*_B}\ , \nonumber\\
D_2 &=& (-1)^{\vp_A} \frac{\im}{\hbar}
\big(J_A+M_{A}\big)J_B
M^{B*}\frac{\delta }{\delta\Phi^*_A}\ -\
(-1)^{\vp_B(\vp_A+1)}\frac{\im}{\hbar}
\big(J_A+M_{A}\big)J_B \frac{\delta  M^{B*}}{\delta\Phi^*_A}\ ,
\nonumber\\
D_3 &=& -\frac{\im}{\hbar} J_B M^{B*} \big(J_A+M_{A}\big)
\frac{\delta }{\delta\Phi^*_A} \ , \nonumber\\
D_4 &=& (-1)^{\vp_B} \Big(\frac{\im}{\hbar}\Big)^2J_BJ_A
M^{A*}M^{B*}\ +\ \frac{\im}{\hbar}J_AM_{\;\;\;B}^{A*}M^{B*}\
,\label{D1234}
\eeq
where the notation
\beq
M_{\;\;\;B}^{A*}\=\frac{\delta^2 M(\Phi,\Phi^*)}{\delta\Phi^*_A\
\delta\Phi^B}
\Big|_{\Phi\rightarrow \frac{\hbar}{\im}\frac{\delta}{\delta J}}
\eeq
was used.

The first term in $D_4$ vanishes identically, whereas the
first one in $D_1$ reads
\beq \label{D1}
(-1)^{\vp_A+1}
\big(J_AJ_B +M_{A}M_{B} + J_AM_{B}+ (-1)^{\vp_A\vp_B}
J_BM_{A} + \sfrac{\hbar}{\im}M_{AB}\big)\frac{\delta
}{\delta\Phi^*_B}\frac{\delta }{\delta\Phi^*_A}\ ,
\eeq
with
\beq
M_{AB} \= \frac{\delta^2 M(\Phi,\Phi^*)
}{\delta\Phi^A\ \delta\Phi^B}\Big|_{\Phi \to
\frac{\hbar}{\im}\frac{\delta}{\delta J}}
\qquad\textrm{so that}\qquad M_{AB}=(-1)^{\vp_A\vp_B}M_{BA}\ .
\eeq
Since under the exchange $A\leftrightarrow B$ the symmetry property
of the expression in brackets is opposite to the symmetry of the
second antifield derivative, (\ref{D1}) vanishes, and $D_1$ is reduced
to the second term.

Next, we collect the remaining terms in~(\ref{D1234}) which are not proportional
to an antifield derivative operator, i.e.~the second terms in $D_2$ and~$D_4$,
\beq \label{A7}
\frac{\im}{\hbar}\Big[J_A\Big(M^{A*}_{\;\;\;B}M^{B*}-M_B\frac{\delta
M^{A*}}{\delta\Phi^*_B}(-1)^{\varepsilon_A}\Big)-
\Big(J_AJ_B+\frac{\hbar}{\im}M_{AB}\Big) \frac{\delta
M^{A*}}{\delta\Phi^*_B}(-1)^{\varepsilon_A}\Big]\ . \label{A2}\eeq
Note that
\beq \frac{\delta M^{A*}}{\delta\Phi^*_B}=\frac{\delta^2
M(\Phi,\Phi^*)}{\delta\Phi^*_B\ \delta\Phi^*_A}\Big|_{\Phi\rightarrow
\frac{\hbar}{\im}\frac{\delta}{\delta J}} \und \frac{\delta
M^{A*}}{\delta\Phi^*_B}=\frac{\delta
M^{B*}}{\delta\Phi^*_A}(-1)^{(\varepsilon_A+1)(\varepsilon_B+1)}\eeq
and, therefore,
\beq \Big(J_AJ_B+\frac{\hbar}{\im}M_{AB}\Big)
\frac{\delta M^{A*}}{\delta\Phi^*_B}(-1)^{\varepsilon_A}=0 \eeq
due to symmetry properties under $A\leftrightarrow B$.
From~(\ref{SoftBrC}), $(M,M)=0$, we have
\beq 0\= \sfrac{1}{2}\frac{\delta }{\delta\Phi^*_A}(M,M)\=
\frac{\delta^2
M(\Phi,\Phi^*)}{\delta\Phi^*_A\ \delta\Phi_B}\frac{\delta
M(\Phi,\Phi^*)}{\delta\Phi^*_B}\ -\ \frac{\delta
M(\Phi,\Phi^*)}{\delta\Phi^B}\frac{\delta^2
M(\Phi,\Phi^*)}{\delta\Phi^*_B\ \delta\Phi^*_A}
(-1)^{\varepsilon_A}\ ,\label{A1}\eeq
which, after substituting $\Phi\rightarrow
\frac{\hbar}{\im}\frac{\delta}{\delta J}$, yields
\beq M^{A*}_{\;\;\;B}M^{B*}\ -\ M_B\frac{\delta
M^{A*}}{\delta\Phi^*_B}(-1)^{\varepsilon_A}\=0 \ .\eeq
We have thus shown that the expression~(\ref{A7}) vanishes.

Finally, the terms in~(\ref{D1234}) proportional to a single antifield
derivative, i.e.~the second term in~$D_1$, the first one in~$D_2$
and all of~$D_3$, have the form \beq
\nonumber && (J_A+M_A)\frac{\delta
M_B}{\delta\Phi^*_A}\frac{\delta}{\delta\Phi^*_B}+
\frac{\im}{\hbar}\big[(-1)^{\vp_A}
(J_A+M_A)J_BM^{B*}-J_BM^{B*}(J_A+M_A)\big]
\frac{\delta}{\delta\Phi^*_A} \\
&&\=\Big(M_A\frac{\delta M_B}{\delta\Phi^*_A}+
M_{BA}M^{A*}(-1)^{\vp_B}\Big)\frac{\delta}{\delta\Phi^*_B}\ .
\label{dPhi*1} \eeq
Again, this expression is equal to zero as a consequence from the analog
of the classical master equation $(M,M)=0$. Indeed,
\beq
0\=\sfrac{1}{2}\frac{\delta }{\delta\Phi^A}(M,M)\= \frac{\delta
M(\Phi,\Phi^*)}{\delta\Phi_B}\frac{\delta^2
M(\Phi,\Phi^*)}{\delta\Phi^*_B\
\delta\Phi^A}+(-1)^{\vp_A}\frac{\delta^2
M(\Phi,\Phi^*)}{\delta\Phi^A\ \delta\Phi^B}\frac{\delta
M(\Phi,\Phi^*)}{\delta\Phi^*_B}\ ,\label{A5}\eeq
thus substituting
$\Phi\rightarrow \frac{\hbar}{\im}\frac{\delta}{\delta J}$
we find
\beq M_{B}\frac{\delta
M_{A}}{\delta\Phi^*_B}\ +\ (-1)^{\varepsilon_A}M_{AB}M^{B*}\=0\ . \eeq
We have proved our assertion that $\hat{q}^2=0$.

\bigskip

\end{document}